\documentclass[12pt]{article}

\usepackage[utf8]{inputenc}
\usepackage{amsmath, amsthm, amssymb, amsfonts}
\usepackage{import}
\usepackage[numbers, comma]{natbib}
\usepackage{fullpage}
\usepackage{authblk}
\usepackage[bibliography=common]{apxproof}
\usepackage{import}
\usepackage{mathtools}
\usepackage{tikz}
\usepackage{float}
\usepackage{enumitem}
\usepackage{bbm}
\usepackage{booktabs}
\usepackage{pdfpages}
\usepackage{graphicx}

\usetikzlibrary{calc,patterns,angles,shapes, positioning, intersections, quotes}

\newtheorem{theorem}{Theorem}
\newtheoremrep{thm}{Theorem}
\newtheoremrep{lem}{Lemma}
\newtheoremrep{clm}{Claim}

\newtheoremrep{prop}{Proposition}

\newtheoremrep{proposition}{Proposition}

\usepackage{graphicx}

\newtheorem{remark}[theorem]{Remark}
\theoremstyle{definition}
\newtheorem{definition}{Definition}
\newtheorem{assumption}{Assumption}
\newtheorem{hypothesis}{Hypothesis}

\usepackage[colorlinks=false,pagebackref=true, hidelinks]{hyperref}

\def\b{\beta}

\newcommand{\ubar}[1]{\text{\underline{$#1$}}}

\def\P{\mathcal{P}}

\def\E{\mathbb{E}}
\newcommand{\df}[1]{\textit{#1}}

\allowdisplaybreaks

\date{\today\\\small{\href{https://goelsumit.com/files/contests_feedback.pdf}{(Link to latest version)}}}

\begin{document}
\title{
Feedback in Dynamic Contests: Theory and Experiment\footnote{The experiment was approved by the IRB at NYU Abu Dhabi and pre-registered with OSF. The pre-registration is available \href{https://osf.io/nzqp8/overview?view_only=19f4c2ce42df43b19d97015c434cea5b}{here}. We gratefully acknowledge financial support from Tamkeen under the NYU Abu Dhabi Research Institute Award CG005.}
}

\author{Sumit Goel\thanks{Division of Social Science, NYU Abu Dhabi; \href{mailto:sumitgoel58@gmail.com}{sumitgoel58@gmail.com}; 0000-0003-3266-9035} \quad Yiqing Yan\thanks{NYU Abu Dhabi; \href{mailto:yy4467@nyu.edu}{yy4467@nyu.edu}} \quad Jeffrey Zeidel\thanks{Center for Behavioral Institutional Design, NYU Abu Dhabi; \href{mailto:jrz8904@nyu.edu}{jrz8904@nyu.edu}}}

\maketitle

\begin{abstract}
We study the effect of interim feedback policies in a dynamic all-pay auction where two players bid over two stages to win a common-value prize. We show that sequential equilibrium outcomes are characterized by Cheapest Signal Equilibria, wherein stage 1 bids are such that one player bids zero while the other chooses a cheapest bid consistent with some signal. Equilibrium payoffs for both players are always zero, and the sum of expected total bids equals the value of the prize. We conduct an experiment with four natural feedback policy treatments--- full, rank, and two cutoff policies---and while the bidding behavior deviates from equilibrium, we fail to reject the hypothesis of no treatment effect on total bids. Further, stage 1 bids induce sunk costs and head starts, and we test for the resulting sunk cost and discouragement effects in stage 2 bidding.
\end{abstract}
\section{Introduction}

Contests are situations in which agents make costly investments to win valuable prizes. In many such settings, investments occur across multiple stages, often accompanied by feedback about the investments of others in earlier stages. For instance, in research and development (R\&D) competitions, firms may learn about the progress of their rivals through public disclosures. In sports or programming tournaments, participants observe their interim standing through leaderboards. In classroom settings, students may receive feedback at intermediate points, such as after a midterm, about the distribution of scores or the proportion of peers who performed above a certain threshold. Across these environments, participants receive various forms of feedback, which may not only inform subsequent investments but also shape initial investments through strategic considerations such as signaling or deterrence. \\

In this paper, we study how feedback policies influence investment behavior in a two-stage all-pay auction model with two players. Each player bids over two stages to win a common-value prize. The player with the higher total bid wins the prize, while both players pay their total bids. Before the auction begins, the auctioneer commits to a feedback policy, defined as a partition of the bid domain such that, after the first stage, each player learns the element of the partition containing their opponent’s stage 1 bid. To ensure existence of equilibrium, we restrict attention to feedback policies that admit a cheapest bid for every signal (i.e., for each element of the partition). \\

We show that sequential equilibrium outcomes are characterized by Cheapest Signal Equilibria (CSE). In equilibrium (on path), stage 1 bidding is such that one player bids zero while the other chooses the cheapest bid corresponding to some signal, and stage 2 bidding then coincides with the (unique) Nash equilibrium of the single-stage all-pay auction with exogenous head starts. We analyze the rank feedback policy separately, as it falls outside the class of policies considered above, and show that both players must bid zero in stage 1 and then mix uniformly between zero and the prize value in stage 2.\\

This equilibrium characterization yields an \df{irrelevance result}: the auctioneer’s equilibrium profit is zero, irrespective of the feedback policy. However, the set of equilibria itself depends on the feedback policy. Apart from the robust equilibrium in which both players bid zero in stage 1, the remaining equilibria are asymmetric and require players to coordinate on which one bids zero in stage 1. When such coordination fails, both players may bid positively in stage 1, thereby generating \df{sunk costs}: costs incurred in stage 1 that yield no strategic benefit to the players in stage 2. Stage 2 bidding may then result in a player’s total bid exceeding the value of the prize, a phenomenon we refer to as the \df{sunk cost effect}. Moreover, stage 2 bidding entails players dropping out with a probability proportional to the perceived difference in stage 1 bids (the \df{head start}), and in general, stage 2 bids are lower when the head start is higher, an instance of the \df{discouragement effect} in our framework. \\

We conduct a laboratory experiment to test these equilibrium predictions of the two-stage all-pay auction model. Our implementation focuses on four natural feedback-policy treatments: full feedback, rank feedback, and two cutoff feedback policies. While we observe overbidding, its magnitude is similar across treatments, and we therefore fail to reject the hypothesis of irrelevance of feedback policies for the auctioneer’s profits. For stage 2 bidding, we observe patterns that are partly consistent with, and partly deviate from, the equilibrium predictions. Higher sunk costs are associated with greater stage 2 bids, in line with the sunk cost fallacy, though the effect is somewhat guarded by the prize value. The discouragement effect of higher head starts kicks in when it is not too small, and is not as large as predicted.  \\

\subsection*{Literature review}

There is a vast literature on all-pay auctions. With multiple players and prizes, \citet*{barut1998symmetric} establish an irrelevance of prize structure for expected equilibrium effort. Our result can be interpreted as extending this irrelevance to feedback policies, albeit in a simple two-player model. \citet*{siegel2014asymmetric} and \citet*{konrad2002investment} characterize equilibria in models with exogenously fixed head starts, which we use in our own equilibrium characterization. Other related work incorporates private abilities (\citet*{moldovanu2001optimal}), arbitrary cost functions (\citet*{fang2020turning}), and noisy output (\citet*{drugov2020tournament}) within the all-pay auction framework. \\

This paper contributes to the literature on feedback design in dynamic contests. Much of the closely related theoretical work in two-stage all-pay auctions assumes noisy output, which guarantees the existence of a Perfect Bayesian Equilibrium (PBE) in pure strategies. With homogeneous agents, \citet*{aoyagi2010information, mihm2019sourcing} identify conditions under which full feedback or no feedback is optimal. Other two-stage models consider asymmetric agents (\citet*{hirata2014model}), private abilities (\citet*{ederer2010feedback}), binary outputs (\citet*{goltsman2011interim}), noisy ranking technologies (\citet*{gershkov2009tournaments}), stage prizes (\citet*{klein2017optimal, sela2012sequential}), pre-contest investments (\citet*{clark2024investing, clark2025investment}), and Tullock contests (\citet*{yildirim2005contests}). Related streams of literature examine feedback policies in elimination contests (\citet*{zhang2009role}), sequential contests (\citet*{hinnosaar2024optimal, deng2023contests}), and continuous time environments (\citet*{ely2023optimal}).\\

There is also a growing empirical and experimental literature on the effect of interim feedback on effort in contests. The work closest to ours is \citet*{ederer2007deception}, who study a two-stage all-pay auction with noisy output and compare biased feedback against full or no feedback. \citet*{fallucchi2013information} highlight the differing effects of feedback in Tullock contests with share and lottery structures. \citet*{dechenaux2023contests} study how information leakage, together with the possibility to revise bids in response, affects behavior in all-pay and Tullock contests. \citet*{azmat2010importance} and \citet*{azmat2019you} examine the role of relative performance feedback on student outcomes in natural field experiments. More recently, \citet*{lemus2021dynamic} and \citet*{hudja2025public} experimentally study the effect of public leaderboards on contest outcomes and obtain contrasting results. For a survey of the experimental literature on contests, see \citet*{dechenaux2015survey}.\\

The paper proceeds as follows. Section 2 presents the model. Section 3 characterizes the sequential equilibrium and discusses the theoretical results. Sections 4 and 5 describe the experimental design and findings, respectively. Section 6 concludes.
 
\section{Model}

\subsection{Two-stage all-pay auction}
Two players $i \in \lbrace 1, 2 \rbrace$, bid across two stages $t \in \lbrace 1, 2 \rbrace$, to win a prize of known common value $v=1$. Denote by $b_{it} \in [0,1]$ the bid of player $i$ in stage $t$, and denote the profile of bids $b = (b_{11},b_{12},b_{21},b_{22})$.\footnote{We restrict bids in any stage to be below 1, because submitting a bid greater than 1 will always yield strictly lower expected utility than bidding 0 (is strictly dominated by 0), and so bids greater than 1 will never occur in equilibrium.} The total bid of player $i$ is $b_{i} = b_{i1} + b_{i2}$. The player with the larger total bid wins the prize, the player with the smaller total bid wins nothing, and both players pay their entire total bid. In case of tie, a fair coin is flipped to determine the winner of the prize. The expected utility of player $i$ as a function of bids is

\begin{align*}
u_{i}(b) = \begin{cases} 1 - b_{i}, & \text{if $b_{i} > b_{-i}$} \\
\frac{1}{2} - b_{i}, & \text{if $b_{i}=b_{-i}$} \\
-b_{i}, & \text{otherwise.} \end{cases}
\end{align*}

\subsection{Feedback policy}
Before the two players participate in this two-stage auction, an auctioneer commits to a \df{feedback policy} $\P$, which is a partition of $[0,1]$. The feedback policy $\P$ determines the information about each player's first stage bid that will be publicly revealed before the players choose their second stage bids. Formally, for any $b_{i1} \in [0,1]$, we define $\P(b_{i1})$ as the element of the partition $\P$ that contains $b_{i1}$, and let it represent the public signal (or message) about player $i$'s first-stage bid. Thus, under feedback policy $\P$, if player $i$ chooses a first-stage bid of $b_{i1}$, then player $j\neq i$ learns that player $i$ chose a first-stage bid in the set $\P(b_{i1})\subset [0,1]$. \\

We will restrict attention to feedback policies which admit a cheapest (smallest) first stage bid for any feasible signal. In other words, the partition $\P$ must be such that each element of $\P$ contains its infimum.\footnote{As we show later, if $\P$ is such that there is an $S\in \P$ with $\inf(S) \notin S$ and $\inf(S)>0$, then no equilibrium exists. Furthermore, implementation in the lab requires a discretization of the bid space, as subjects can only be paid in discrete amounts, so any implemented policy implicitly satisfies this assumption.} 

\begin{assumption}
\label{ass:cheapest_signal_feedback}
The feedback policy $\P$, which is a partition of $[0,1]$, is such that for all $S\in \P$, $$\ubar{S} =\inf(S) \in S.$$
\end{assumption}

We now present some examples of feedback policies:

\begin{enumerate}
    \item No feedback: This policy reveals no information about the first stage bids, and is captured by
    $$\P^{NONE}=\{[0,1]\}.$$
    \item Full feedback: This policy reveals exactly the first stage bids, and is captured by
    $$\P^{FULL}=\{\{b_{i1}\}:b_{i1} \in [0,1]\}.$$
    \item Cutoff feedback: This policy reveals if a player bid at least $c$ or strictly less than $c$, and is captured by
    \begin{align*}
\P^{CUTOFF}(c)= \{[0,c), [c, 1]\}.
\end{align*}
\end{enumerate}

While our definition precludes some natural policies, such as one which reveals how the players rank in terms of their first stage bids, we will discuss later how our results extend to such policies.

\subsection{Sequential Equilibrium}

Given any feedback policy $\P$, the two-stage all-pay auction defines an extensive-form game with imperfect information between the two players. Formally, the game proceeds as follows:

\begin{enumerate}
    \item Both players simultaneously choose their first stage bids, $b_{11}, b_{21} \in [0,1]$.
    \item The feedback policy $\P$ generates public signals $\P(b_{11}), \P(b_{21}) \in \P$.
    \item Both players simultaneously choose their second stage bids, $b_{12}, b_{22} \in [0,1]$.
    \item Payoffs are realized as per the rules of the two-stage all-pay auction.
\end{enumerate}

\begin{remark}
This extensive-form game can be more formally described as one in which player 1 bids first in both stages, and player 2 bids second in both stages with no additional information about the corresponding stage bid of player 1. For expositional purposes we define the game as above and define corresponding solution concepts. An analysis of the more formal extensive form yields the same theoretical results. 

\end{remark}

We first introduce the solution concept of Perfect Bayesian Equilibrium (PBE). This constitutes describing for each player, their bidding strategy in the two stages as well as their belief about the first stage bid of their opponent given the feedback, so that the beliefs are consistent with the bidding strategies, and the bidding strategies are rational given the beliefs and opponent's strategy. We now formalize this notion.\\

A \df{strategy} for player $i$ is a pair $\b_{i}=(\b_{i1}, \b_{i2})$, where:
\begin{itemize}
    \item $\b_{i1}\in \Delta [0,1]$ is a probability distribution over first stage bids.
    \item $\b_{i2}: [0,1]\times \P \to \Delta [0,1]$ specifies a probability distribution over second stage bids, given the player’s own first-stage bid and the observed feedback signal.
\end{itemize}

A \df{belief} for player $i$ is a mapping $\mu_i:[0,1]\times  \P \to \Delta [0,1]$ that specifies player $i$'s belief about their opponent's first stage bid, given their own first stage bid and the observed feedback signal. An \df{assessment} is a pair $(\b, \mu)$, where $\b=(\b_1, \b_2)$ is a strategy profile and $\mu=(\mu_1, \mu_2)$ is a belief profile. \\

An assessment $(\b, \mu)$ is a \df{perfect Bayesian equilibrium (PBE)} if for each player $i\in \{1, 2\}$:
\begin{itemize}
    \item Beliefs are Bayesian: The belief $\mu_i$ is Bayes consistent with strategy profile $\b$, meaning beliefs are updated using Bayes' rule whenever possible.
    \item Strategies are sequentially rational: The strategy $\b_i$ maximizes expected utility of player $i$, given $\b_{-i}$ and $\mu_i$, at every decision node.
\end{itemize}

We will focus on the sequential equilibrium of this game, which is a refinement of PBE that imposes constraints on beliefs at information sets that are reached with probability $0$. Formally, an assessment $(\b, \mu)$ is a \df{sequential equilibrium} if it is a PBE, and there exists a sequence of completely mixed strategy profiles $\beta^1, \beta^2, \dots, $ such that $\lim_{n\to \infty } \beta^n = \beta$ and $\lim_{n\to \infty } \mu^n = \mu$, where $\mu^n$ is the unique Bayesian belief profile under strategy profile $\beta^n$.\\

In a PBE, a player's belief $\mu_{i}$ about their opponent's first stage bid can, absurdly, depend on their own first stage bid. Sequential equilibrium requires beliefs to depend only on the observed feedback.

\begin{lemrep}
\label{lem:sequential_belief_restriction}
If $(\b, \mu)$ is a sequential equilibrium, then $\mu_i$ must be such that for any $b_{i1}, b'_{i1} \in [0,1]$ and $S\in \P$, 
$$\mu_i(b_{i1}, S) = \mu_i(b_{i1}', S).$$
\end{lemrep}
\begin{proof}
Fix any $i\in \{1, 2\}$, $b_{i1}, b'_{i1}\in [0,1]$ and $S\in \P$. 
Since $(\b, \mu)$ is a sequential equilibrium, there exists a sequence of completely mixed strategy profiles $\beta^1, \beta^2, \dots, $ such that $\lim_{n\to \infty } \beta^n = \beta$ and $\lim_{n\to \infty } \mu^n = \mu$, where $\mu^n$ is the unique Bayesian belief profile under strategy profile $\beta^n$. It follows that for each $n$, 
$$\mu_i^n(b_{i1}, S) = \mu_i^n(b_{i1}', S),$$
and hence, it must be that $\mu_i(b_{i1}, S) = \mu_i(b_{i1}', S)$.

\end{proof}

With this, if $(\b, \mu)$ is a sequential equilibrium, we simply write $\mu_i:\P \to \Delta [0, 1]$. Further, we say $(\b, \mu)$ is a \df{pure-strategy sequential equilibrium} if $\b_{i1}=\delta_{b_{i1}}$ for $i\in \{1, 2\}$, where $\delta_x$ denotes the Dirac measure at $x\in [0,1]$.

\subsection{Total equilibrium bids}
We are interested in examining how feedback policies influence the auctioneer's profit, as well as the equilibrium payoffs of the two players. Given a feedback policy $\P$ and an assessment $(\b, \mu)$ of the induced game, let $b_i\sim F_i$ denote the (random) total bid of player $i$. Then, the expected payoff of player $i$ is
$$u_i(\b, \mu) = \Pr[b_i>b_{-i}]+\frac{1}{2}\Pr[b_i=b_{-i}]-\E[b_i],$$
and the auctioneer's expected profit is
$$\pi(\b, \mu) = \E[b_1+b_2]-1.$$
Notice that, by definition, the payoffs and profit must be such that $$u_1(\b, \mu)+u_2(\b, \mu)+\pi(\b, \mu)=0.$$
Our goal is to investigate how feedback policies influence each of these quantities in equilibrium. 
\section{Theoretical Results}
In this section, we state our main irrelevance result and discuss some important equilibrium properties that drive the result.

\subsection{Irrelevance result}

Before stating our result, we recall the classical all-pay auction model where instead of bidding over two stages, the players submit a single bid. In this normal-form game, there is a unique Nash equilibrium, and it is such that both players bid uniformly on the interval $[0,1]$. Consequently, the equilibrium payoff of both players is zero, and the auctioneer's profit is also zero (\citet*{barut1998symmetric}).\\

Our main result is that these properties of the equilibrium extend to our two-stage all-pay auction model, irrespective of the feedback policy in place.

\begin{thm}
\label{thm:irrelevance}
For any feedback policy $\P$ that satisfies Assumption \ref{ass:cheapest_signal_feedback}, and any pure strategy sequential equilibrium $(\b, \mu)$, 
$$u_1(\b, \mu)=0, u_2(\b, \mu)=0, \text{ and } \pi(\b, \mu) =0.$$
\end{thm}

To prove this result, we will characterize pure-strategy sequential equilibria under any feedback policy $\P$, and show that an equilibrium $(\b, \mu)$ always exhibits the above properties.

\subsection{Equilibrium characterization}

Consider any feedback policy $\P$ that satisfies Assumption \ref{ass:cheapest_signal_feedback}. Intuitively, since bids in the two stages are perfect substitutes, the only potential benefit from choosing a non-zero first stage bid arises from the signal it generates. Given this, it is reasonable to suspect that, for any target signal, a player should choose the cheapest possible first stage bid which generates the signal. Based on this, we define the corresponding cheapest signal belief for the players.

\begin{definition}
A belief $\mu_i: \P \to \Delta [0,1]$ is a \df{Cheapest Signal Belief (CSB)} if, for any signal $S\in \P$, 
$$\mu_i(S)=\delta_{\ubar{S}}.$$

\end{definition}

In other words, under the CSB, whenever a player observes a public signal $S\in \P$, their belief about their opponent's first stage bid is entirely concentrated at the lowest possible bid that could have generated $S$, namely $\ubar{S}$. \\

We now define a Cheapest Signal Strategy (CSS), which not only ensures that a player chooses the cheapest possible first-stage bid associated with some signal, but also specifies their second-stage bidding behavior under the CSB. The definition is inspired by the unique equilibrium of a single-stage all-pay auction with exogenous head starts (\citet*{siegel2014asymmetric, konrad2002investment}), which we restate below in our framework using the context of the Full feedback policy.

\begin{lem}
\label{lem:one_shot_all_pay}
Consider the feedback policy $\P^{FULL}$. For any first-stage bids $b_{11}, b_{21}\in [0,1]$, there exists a unique Nash equilibrium in second-stage bids. In this equilibrium, for each player $i\in \{1, 2\}$, the total bid of player $i$ is

$$
b_{i1} + b_{i2} = \begin{cases} b_{i1} & \text { with probability } |b_{11}-b_{21}| \\ U(\max\{b_{11}, b_{21}\}, \min\{b_{11}, b_{21}\}+1] & \text { otherwise }\end{cases}.
$$
Furthermore, each player’s equilibrium utility, accounting for the cost of first-stage bids, is
$$u_i=-\min\{b_{11}, b_{21}\}.$$
\end{lem}

Thus, each player bids zero in the second stage with a probability equal to the absolute difference in first-stage bids (i.e., the head start), and otherwise ensures that their total bid is uniformly distributed on the interval $(\max\{b_{11}, b_{21}\}, \min\{b_{11}, b_{21}\}+1]$.\\

Returning to the definition of CSS, given any $b_{i1}\in [0,1]$ and signal $S_{-i} \in \P$, a player holding the CSB believes that the winning total bid must be at least $\max \{b_{i1}, \ubar{S_{-i}}\}$. At the same time, they believe that some player's total bid cannot exceed $\min \{b_{i1},\ubar{S_{-i}}\} + 1$. Hence, in the second stage, player $i$ will either bid $0$ (so that their total bid equals their first-stage bid) or bid positively and ensure their total bid lies within the interval, $(\max \lbrace b_{i1},\ubar{S_{-i}} \rbrace, \min \lbrace b_{i1}, \ubar{S_{-i}} \rbrace + 1]$. In the CSS, we specify the exact distribution of bids for some important information sets, drawing on the structure in Lemma \ref{lem:one_shot_all_pay}.

\begin{definition}
A strategy $\b_i =(\b_{i1}, \b_{i2})$ is a \df{Cheapest Signal Strategy (CSS)} if:
\begin{enumerate}
    \item There exists an $S_i\in \P$ such that $\b_{i1}=\delta_{\ubar{S_i}}.$
    \item Fix any $b_{i1} \in [0,1]$ and $S_{-i}\in \P$. 
    \begin{enumerate}
        \item If $b_{i1}=\ubar{\P(b_{i1})},$ then 
        $\b_{i2}(b_{i1}, S_{-i})$ is such that
$$
b_{i1} + b_{i2} = \begin{cases} b_{i1} & \text { with probability } |b_{i1}-\ubar{S_{-i}}| \\ U(\max\{b_{i1}, \ubar{S_{-i}}\}, \min\{b_{i1}, \ubar{S_{-i}}\}+1] & \text { otherwise }\end{cases}.
$$
\item Otherwise, $\b_{i2}(b_{i1}, S_{-i})$ is such that $$
b_{i1} + b_{i2} \in \Delta (\max\{b_{i1}, \ubar{S_{-i}}\}, \min\{\ubar{\P(b_{i1})}, \ubar{S_{-i}}\}+1].
$$
    \end{enumerate}
\end{enumerate}

\end{definition}

Under CSS, in the first stage, a player chooses the cheapest possible first-stage bid associated with some signal. Furthermore, for the second stage, on information sets where it did choose a cheapest signal in the first stage, it's bidding is as prescribed by the second-stage equilibrium strategy in Lemma \ref{lem:one_shot_all_pay}, with $b_{-i1}$ replaced by $\ubar{S_{-i}}$. On information sets where it did not chose the cheapest signal, CSS only requires that the total bid is in the interval $(\max\{b_{i1}, \ubar{S_{-i}}\}, \min\{\ubar{\P(b_{i1})}, \ubar{S_{-i}}\}+1]$, with no restriction on the distribution.\\

We are now ready to present our characterization of pure-strategy equilibrium outcomes. We first show that any assessment $(\b, \mu)$ in which both agents play CSS and hold CSB, and at least one agent bids $0$ in stage 1, constitutes a sequential equilibrium. The beliefs are clearly consistent under CSS, and the close correspondence of second-stage bids in CSS to those in Lemma \ref{lem:one_shot_all_pay} ensures that the strategies are sequentially rational. Moreover, at least one agent must bid $0$ in stage 1. Intuitively, if both agents were to bid positive amounts in the first stage, sequential rationality would imply that on-path second-stage bids follow Lemma \ref{lem:one_shot_all_pay}, yielding negative equilibrium utilities. An agent could then profitably deviate by bidding $0$ in both stages, thereby obtaining a strictly higher payoff.

\begin{proprep}
\label{prop:CSS_is_equilibrium}
Any assessment $(\b,\mu)$ in which $\b_1$ and $\b_2$ are Cheapest Signal Strategies with $\b_{i1}=\delta_0$ for some $i\in \{1,2\}$, and $\mu_1,\mu_2$ are Cheapest Signal Beliefs, is a sequential equilibrium.  
\end{proprep}

\begin{proof}
Suppose $(\b, \mu)$ is such an assessment. We will show that $(\b, \mu)$ is a sequential equilibrium. 

\begin{enumerate}
    \item Beliefs are Bayesian: For player $i$, since $\b_{-i}$ is a CSS, the belief $\mu_i$ is indeed Bayesian consistent. 
    \item Strategies are sequentially rational: 
    
    \textbf{Stage 2}
    
    Consider player $i$ and suppose it bid $b_{i1}$ and received a signal $S_{-i}\in \P$. Then, $\mu_i=\delta_{\ubar{S_{-i}}}$. Given $\b_{-i}$, its belief about the total bid of the other player will be 
    $$
b_{-i} = \begin{cases} \ubar{S_{-i}} & \text { with probability } |\ubar{S_{-i}}-\ubar{\P(b_{i1})}| \\ U(\max\{\ubar{S_{-i}}, \ubar{\P(b_{i1})}\}, \min\{\ubar{S_{-i}}, \ubar{\P(b_{i1})}\}+1] & \text { otherwise }\end{cases}.
$$

From here, it is clear that an optimal $b_{i2}$ must be such that $$b_i\in \{b_{i1}\} \cup (\max\{\ubar{S_{-i}}, b_{i1}\}, \min\{\ubar{S_{-i}}, \ubar{\P(b_{i1})}\}+1].$$
Moreover, within the interval $(\max\{\ubar{S_{-i}}, b_{i1}\}, \min\{\ubar{S_{-i}}, \ubar{\P(b_{i1})}\}+1]$, the marginal benefit from increasing $b_i$ is equal to the marginal cost, and hence the player must be indifferent between all such bids. Thus, to find the optimal bid(s), we can simply compare the utility from choosing $b_i=b_{i1}$ (i.e., $b_{i2}=0$) to that from choosing $b_i=\min\{\ubar{S_{-i}}, \ubar{\P(b_{i1})}\}+1$.

\begin{enumerate}
    \item $b_i=b_{i1}$: In this case, player $i$'s utility equals $\Pr[b_{-i}<b_{i1}]$. If $b_{i1}>\ubar{S_{-i}}$, then this probability equals $b_{i1} - \min\{\ubar{S_{-i}}, \ubar{\P(b_{i1})}\}$. If $b_{i1}\leq \ubar{S_{-i}}$, then this probability equals $0$.
    \item $b_i=\min\{\ubar{S_{-i}}, \ubar{S_i'}\}+1$: In this case, player $i$ wins for sure, and hence it's utility is $$1-b_{i2}=b_{i1} - \min\{\ubar{S_{-i}}, \ubar{\P(b_{i1})}\}. $$
\end{enumerate}

    It follows that $\b_{i2}(b_{i1}, S_{-i})$, as defined under the CSS, is optimal.

    \textbf{Stage 1}
    
    Consider player $i$. Given $\b_{-i}$, it follows from above that the gain from bidding $b_{i1}$ in stage $1$ is 
    $$b_{i1} - \min\{\ubar{S_{-i}}, \ubar{\P(b_{i1})}\},$$
    while the cost is $b_{i1}$. Thus, the utility is simply 
    $$-\min\{\ubar{S_{-i}}, \ubar{\P(b_{i1})}\}.$$
    Now if $\ubar{S_{-i}}\neq 0$, $b_{i1}=0$ is optimal. And if $\ubar{S_{-i}}= 0$, any $b_{i1}$ such that $b_{i1}=\ubar{\P(b_{i1})}$ is optimal. It follows that $\b_1$ and $\b_2$ are sequentially rational.

\item Beliefs are consistent: It is straightforward to construct a sequence of completely mixed strategies that converge to $\b_1, \b_2$ so that the induced Bayesian beliefs converge to $\mu_1, \mu_2$. The main restriction introduced is that the belief $\mu_i$ must depend only on the feedback $S_{-i}\in \P$ (Lemma \ref{lem:sequential_belief_restriction}), which is indeed the case.
\end{enumerate}
Thus, $(\b, \mu)$ is a sequential equilibrium.
\end{proof}

We will refer to the set of all such assessments as Cheapest Signal Equilibria (CSE). \\

To complete the characterization, we next show that any pure-strategy sequential equilibrium must be outcome-equivalent to a Cheapest Signal Equilibrium. As before, in any candidate pure-strategy equilibrium, sequential rationality implies that on-path second-stage bids follow the structure in Lemma \ref{lem:one_shot_all_pay}. Thus, for the same reason discussed above, at least one agent must bid $0$ in the first stage. Furthermore, the other agent must also be choosing the cheapest first-stage bid consistent with some signal. If not, we show that this agent can deviate to a smaller first-stage bid that induces the same signal and obtain strictly higher utility.\footnote{The restriction to sequential equilibrium is critical for this argument. The reasoning does not extend to PBE: under a PBE, a player may revise their belief about the opponent’s first-stage bid when deviating to a smaller bid, potentially eliminating the profitable deviation. As we show in Lemma \ref{lem:sequential_belief_restriction}, such a revision of beliefs is not possible under sequential equilibrium.} Hence, every pure-strategy sequential equilibrium is outcome-equivalent to a Cheapest Signal Equilibrium in which agents choose the corresponding first-stage bids.

\begin{proprep}
\label{prop:CSS_is_wlog}
Every pure-strategy sequential equilibrium is outcome-equivalent to some Cheapest Signal Equilibrium (i.e., the induced on-path bidding is identical). 
\end{proprep}
\begin{proof}
Suppose $(\b, \mu)$ is a pure-strategy sequential equilibrium. Then, for each $i\in \{1, 2\}$, there is a $b_{i1}\in [0,1]$ such that $\b_{i1}=\delta_{b_{i1}}$. Since the belief should be Bayesian and consistent, we must have 
$$\mu_{i}(\P(b_{-i1})) = \delta_{b_{-i1}}.$$
Sequential rationality further implies that, on path, $\b_{i2}(b_{i1}, \P(b_{-i1}))$ is such that 

    $$
b_{i} = \begin{cases} b_{i1} & \text { with probability } |b_{11}-b_{21}| \\ U(\max\{b_{11}, b_{21}\}, \min\{b_{11}, b_{21}\}+1] & \text { otherwise }\end{cases},
$$

and the equilibrium utility of player $i$ is $-min\{b_{11}, b_{21}\}$ (Lemma \ref{lem:one_shot_all_pay}). \\

Now if both $b_{11}, b_{21}>0$, a player can deviate to bidding $0$ in both stages and obtain a strictly higher utility, which contradicts $(\b, \mu)$ being an equilibrium. Thus, there must be some $j\in \{1, 2\}$ such that $b_{j1}=0$. Now consider $i\neq j$. Given $\b_{-i}$ as above, observe that if there exists a $b'_{i1}\in \P(b_{i1})$ such that $0<b'_{i1}<b_{i1}$, then player $i$ can bid $b'_{i1}$ in stage 1, and bid $b'_{i2}=0$ in stage 2, obtaining a utility of $b_{i1}-b'_{i1}>0$, which again contradicts $(\b, \mu)$ being a sequential equilibrium. Thus, it must be that $b_{i1}=\ubar{\P(b_{i1})}$. This equilibrium $(\b, \mu)$ is outcome equivalent to the Cheapest Signal Equilibrium $(\b', \mu')$ where $\b_{j1}'=\delta_0$ and $\b'_{i1}=\delta_{\ubar{S}}$ where $S=\P(b_{i1})$. 
\end{proof}

It follows that for the purpose of analyzing equilibrium outcomes, it is without loss of generality to restrict attention to Cheapest Signal Equilibria. We now prove Theorem \ref{thm:irrelevance}.

\begin{proof}[Proof of Theorem \ref{thm:irrelevance}]
Suppose feedback policy $\P$ satisfies Assumption \ref{ass:cheapest_signal_feedback} and $(\b, \mu)$ is a pure-strategy sequential equilibrium under $\P$. By Proposition \ref{prop:CSS_is_wlog}, there exists a Cheapest Signal Equilibrium $(\b', \mu')$ which is outcome-equivalent. In this equilibrium, observe that there is some $i\in \{1, 2\}$ such that $0$ is in the support of $b_i$, and $\Pr[b_{-i}\leq 0]=0$. It follows that $u_i(\b', \mu')=0$. For $j\neq i$, we have that $1$ is in the support of $b_j$ and $\Pr[b_{-j}<1]=1$. It follows that $u_j(\b', \mu')=0$. As a consequence, we also get that the auctioneer's expected profit is $\pi(\b', \mu')=0$. By outcome equivalence, these properties extend to the equilibrium $(\b, \mu)$.  
\end{proof}

\subsection{Rank feedback}
Theorem \ref{thm:irrelevance} establishes irrelevance for a fairly broad family of feedback policies. However, it excludes some natural policies, such as one where the players are ranked based on their first stage bids (with ties broken uniformly at random). We refer to this policy as \df{Rank feedback} ($\P^{RANK}$): if player $i$ bids $b_{i1}$ in stage $1$, it learns whether $b_{-i1}\in [0, b_{i1}]$ or $b_{-i1}\in[b_{i1}, 1]$. We show that in any equilibrium under Rank feedback, both players bid $0$ in stage 1, and then play the one-shot equilibrium of bidding $U[0,1]$ in stage 2 (on path). It follows that Theorem \ref{thm:irrelevance} extends to the Rank feedback policy.

\begin{proprep}
\label{prop:rank}
Consider the feedback policy $\P^{RANK}$. If $(\b, \mu)$ is a pure-strategy sequential equilibrium, it must be that $\b_{11}=\b_{21}=\delta_0$ and $b_i\sim U(0,1]$ (on path). Moreover, such an equilibrium exists.
\end{proprep}
\begin{proof}
Suppose $(\b, \mu)$ is a pure-strategy sequential equilibrium. Then, for each $i\in \{1, 2\}$, there is a $b_{i1}\in [0,1]$ such that $\b_{i1}=\delta_{b_{i1}}$. Using an argument analogous to that of Proposition \ref{prop:CSS_is_wlog}, we can show that there must be some $j\in \{1, 2\}$ such that $b_{j1}=0$. Further, for $i\neq j$, if $b_{i1}>0$, then player $i$ can deviate to bidding $b'_{i1}$ such that $0<b'_{i1}<b_{i1}$ in stage 1 and $b'_{i2}=0$ in stage 2, and obtain a strictly higher utility. Thus, it must be that $\b_{11}=\b_{21}=\delta_0$, and hence, $b_i\sim U(0,1]$ on path.\\

We now show by construction that such an equilibrium exists. The construction $(\b, \mu)$ is a simple adaptation of CSS and CSB for the rank feedback policy, defined as follows: For each $i\in \{1, 2\}$, let $$\mu_i(b_{i1}, [0, b_{i1}])=\delta_0 \text{ and } \mu_i(b_{i1}, [b_{i1}, 1])=\delta_{b_{i1}},$$

and let $\b_i$ be defined as follows:
\begin{enumerate}
    \item $\b_{i1}=\delta_0$,
    \item For any $b_{i1}\in [0,1]$,
    \begin{enumerate}
        \item If $S_{-i}=[0, b_{i1}]$, then $\b_{i2}(b_{i1}, S_{-i})$ is such that    
        $$
b_{i} = \begin{cases} b_{i1} & \text { with probability } b_{i1} \\ U(b_{i1},1] & \text { otherwise }\end{cases},
$$
\item If $S_{-i}=[b_{i1}, 1]$, then $\b_{i2}(b_{i1}, S_{-i})$ is such that    
$$b_{i}\sim U(b_{i1}, 1]. $$
    \end{enumerate}
\end{enumerate}

It is straightforward to verify that the strategies are sequentially rational, and the beliefs are Bayesian and consistent. 
\end{proof}

\subsection{Second stage bidding}
Given any feedback policy $\P$, it follows from Propositions \ref{prop:CSS_is_equilibrium} and \ref{prop:CSS_is_wlog} that any pure-strategy sequential equilibrium involves at least one player bidding $0$ in stage 1. Consequently, any such equilibrium, apart from the one where both players bid $0$ in stage 1, is asymmetric and essentially require the players to coordinate on which player bids $0$ in the first stage. However, it is reasonable to suspect that coordination failures may occur, and both players enter stage 2 having bid positively in stage 1. In such cases, the stage 1 bids are sunk costs, and bidding behavior in stage 2 should treat them as such. We note some testable implications for second stage bidding under Cheapest Signal Strategies in the following proposition. 

\begin{proprep}
\label{prop:CSS_second_stage}
If $\b_i$ is a Cheapest Signal Strategy with $\b_{i1}=\delta_{b_{i1}}$, then for any $S_{-i}\in \P$, we have the following:

\begin{enumerate}
   \item $\Pr[b_{i1}+b_{i2}>1]=\min\{b_{i1}, \ubar{S_{-i}}\}$ (the \df{sunk cost}).
    \item $\Pr[b_{i2}=0]=|b_{i1}-\ubar{S_{-i}}|$ (the \df{head start}).
 
    \item $\E[b_{i2}]= \begin{cases} \frac{1}{2}\left[1-(b_{i1}-\ubar{S_{-i}})\right]^2 & \text{if } b_{i1} \geq  \ubar{S_{-i}} \\ \frac{1}{2}\left[1-(b_{i1}-\ubar{S_{-i}})^2\right] & \text{if } b_{i1} \leq  \ubar{S_{-i}} .\end{cases}
$
\end{enumerate}

\end{proprep}

\begin{proof}
The first two claims follow directly from the definition of CSS. For the third claim, first suppose $b_{i1}\geq \ubar{S_{-i}}$. Then, by definition,
\begin{align*}
    \E[b_{i2}]&=0*|b_{i1}-\ubar{S_{-i}}|+\frac{(0+\ubar{S_{-i}}+1-b_{i1})}{2}(1-|b_{i1}-\ubar{S_{-i}}|)\\
    &=\frac{1}{2}\left[1-(b_{i1}-\ubar{S_{-i}})\right]^2.
\end{align*}

Similarly, if $b_{i1}\leq \ubar{S_{-i}}$,

\begin{align*}
    \E[b_{i2}]&=0*|b_{i1}-\ubar{S_{-i}}|+\frac{(\ubar{S_{-i}} - b_{i1}+1)}{2}(1-|b_{i1}-\ubar{S_{-i}}|)\\
    &=\frac{1}{2}\left[1-(b_{i1}-\ubar{S_{-i}})^2\right].
\end{align*}
\end{proof}

In words, if player $i$ is playing CSS and bids $b_{i1}$ in stage 1, and observes signal $S_{-i}$, then their second stage bidding should be such that their total bid exceeds the value of the prize with a probability equal to $\min\{b_{i1}, \ubar{S_{-i}}\}$. Intuitively, $\min\{b_{i1}, \ubar{S_{-i}}\}$ is player $i$'s sunk cost and shouldn't affect their second stage bids. This leads to the \df{sunk cost effect}: if player $i$'s sunk costs increase, so should the probability that their total bid exceeds the value of the prize. Further, player $i$ should bid $0$ (drop out) in the second stage with a probability equal to the perceived head start, $|b_{i1}-\ubar{S_{-i}}|$, and more generally, the mean second stage bid is decreasing in this head start. This presents an instance in our framework of the classical \df{discouragement effect}: if the perceived head start increases, player $i$'s second stage bids should be lower. 
\section{Experiment}

In this section, we present a laboratory experiment designed to test the key equilibrium predictions of our two-stage all-pay auction model: the irrelevance of feedback policy for auctioneer's profit (Theorem \ref{thm:irrelevance} and Proposition \ref{prop:rank}), and the sunk cost effect and the discouragement effect in second stage bidding behavior under Cheapest Signal Strategies (Proposition \ref{prop:CSS_second_stage}). 

\subsection{Treatments}

In our implementation of the two-stage all-pay auction, each subject was endowed with a starting `balance' of 40 dirhams, and submitted (integer) bids in two stages for a prize worth 20 dirhams. In the first stage, bids could be any amount from 0 to 40. In the second stage, bids were constrained so that the total bid would not exceed the starting balance: if the first stage bid was $k$, the second stage bid must be between $0$ to $40-k$. \\

After the first stage, subjects were provided feedback about the first stage bid of their opponent according to a feedback policy, which was publicly announced at the beginning. After second stage, the exact bids of both players in the two stages were revealed. The player with the higher total bid received the prize, and both players paid their total bid from their starting balance and kept the remainder. The final payoff, or the `final balance', was 40 + 20 - total bid for the high bidder and 40 - total bid for the low bidder. All ties were broken by a fair virtual coin flip.\\   

Subjects participated in four feedback policy treatments:
\begin{enumerate}
    \item Full feedback (F): Under this policy, subjects were informed about the exact stage 1 bid of their opponent.
    \item Cutoff 5 feedback (C5): Under this policy, subjects were informed whether or not their opponent's stage 1 bid was at least $5$.
    \item Cutoff 10 feedback (C10): Under this policy, subjects were informed whether or not their opponent's stage 1 bid was at least $10$.
    \item Rank feedback (R): Under this policy, subjects were informed whether their opponent's stage 1 bid was higher or lower than their own stage 1 bid, with ties again broken by a fair virtual coin flip. Consequently, each subject could infer whether they were the \df{leader} or the \df{laggard} at the end of stage 1.
\end{enumerate}

For our analysis, we normalize the data by dividing all bids and cutoffs by 20 so as to maintain consistency with the theoretical model.

\subsection{Procedures}

We ran 12 sessions at the SSEL laboratory at the NYU Abu Dhabi campus. Subjects were recruited from the undergraduate student population of NYU Abu Dhabi. Each session had between 10 and 20 subjects, and lasted about one hour.  

\begin{table}[!h]
\centering
\caption{Session Information}
\centering
\begin{tabular}[t]{lrrr}
\toprule
Session & Date & Treatment Order & Subjects \\
\midrule
1 & 11/28/24  & F, R, C5, C10  & 12  \\
2 & 1/27/25  & R, C10, C5, F  & 14  \\
3 & 1/31/25 & C5, C10, F, R  & 20  \\ 
4 & 2/22/25 & R, C5, F, C10  & 10  \\
5 & 5/23/25  & F, C10, R, C5 & 12  \\
6 & 11/28/24 & R, F, C10, C5 & 16  \\ 
7 & 11/29/24  & C5, F, R, C10  & 18  \\
8 & 11/29/24  & C10, F, C5, R & 14 \\ 
9 & 12/06/24 & F, C5, C10, R  & 16  \\
10 & 12/13/24 & C10, C5, R, F  & 14  \\
11 & 1/24/25 & C5, R, C10, F  & 16  \\
12 & 1/24/25 & C10, R, F, C5  & 18 \\ 
\bottomrule
\end{tabular}
\end{table}

At the beginning of each session, subjects read through instructions and answered quiz questions, and could only advance to the game after answering all quiz questions correctly. After all subjects completed the quiz, five training rounds of the two-stage auction game with a no information feedback policy were played, to familiarize subjects with the interface and the game. Five rounds of each of the four feedback policy treatments were then played in turn, so that each of the 180 subjects played a total of 20 games. The five rounds of a treatment were played back-to-back, to make it easier for subjects to understand and remember the feedback policy that applied in a given round. Subjects were randomly re-matched between all rounds, including the no information practice rounds. The order of the treatments was shuffled between sessions, so that each treatment appeared first, second, third or fourth in the treatment order exactly three times. This was done to eliminate treatment order effects in the statistical analysis. The experimental interface and instructions are provided in Appendix \ref{sec:experiment_instructions}.

\AtEndDocument{
  \clearpage
  \section{Experimental Instructions}
  \label{sec:experiment_instructions}
  \addcontentsline{toc}{section}{Experimental Instructions}
  \includepdf[pages=-]{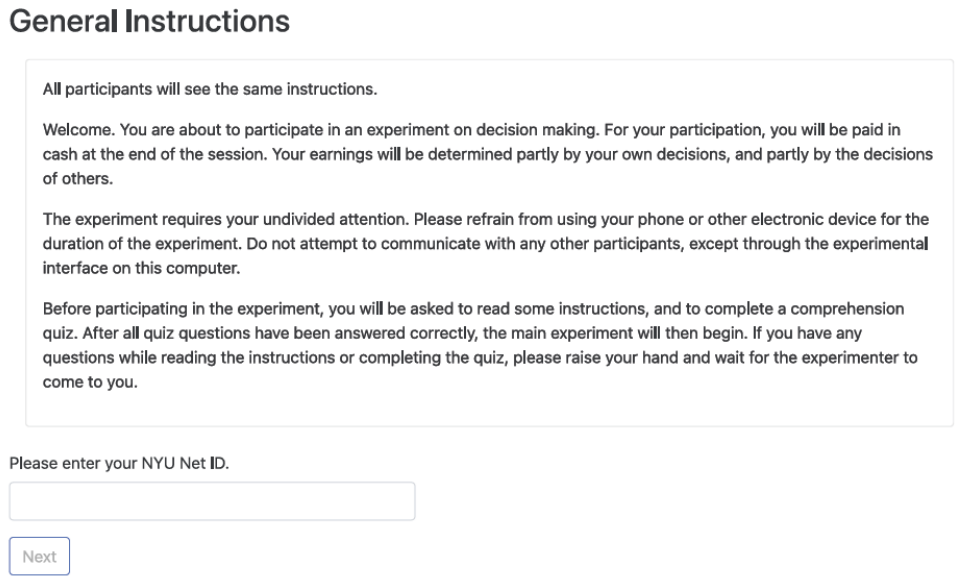}
}

\subsection{Hypotheses}

We present five hypotheses based on our theoretical results. The first two are derived from the irrelevance result in Theorem \ref{thm:irrelevance} and Proposition \ref{prop:rank}.

\begin{hypothesis}[Irrelevance - 1]
\label{hypo:same_profit}
The average profit of the auctioneer is the same in every treatment.
\end{hypothesis}

Hypothesis \ref{hypo:same_profit} is the main hypothesis of interest from the design perspective. It says that the auctioneer in the two-stage all-pay auction cannot influence profits with their choice of feedback policy. 

\begin{hypothesis}[Irrelevance - 2]
\label{hypo:zero_profit}

The average profit of the auctioneer is zero in every treatment. 
\end{hypothesis}

Hypothesis \ref{hypo:zero_profit} is stronger than Hypothesis \ref{hypo:same_profit}, in that it says the average profit is not just the same, but zero across treatments. In particular, if we observe over-bidding relative to equilibrium but this overbidding is equal across feedback policies, then we may reject Hypothesis \ref{hypo:zero_profit} while at the same time be unable to reject Hypothesis \ref{hypo:same_profit}.\\

The remaining three hypotheses are about the sunk cost effect and discouragement effect in second stage bidding behavior identified in Proposition \ref{prop:CSS_second_stage}. Since the result assumes that the player is playing Cheapest Signal Strategies, we state the hypotheses under the following assumption.

\begin{assumption}
\label{ass:CSS_play}
The first stage bid of player $i$, $b_{i1}$, is the cheapest bid that generates the signal $\P(b_{i1})$ (i.e., $b_{i1}=\ubar{\P(b_{i1})}$).
\end{assumption}

When Assumption \ref{ass:CSS_play} does not hold, the definition of CSS is flexible and does not put much structure on what the second stage bidding behavior should be. In fact, in such a case, there is evidence that the player is not playing CSS, or for that matter, any pure strategy that could be sustained in pure-strategy sequential equilibrium. Therefore, we will state and test the following hypotheses only under Assumption \ref{ass:CSS_play}.

\begin{hypothesis}[Sunk cost effect]
\label{hypo:exceed_prize}
Under Assumption \ref{ass:CSS_play}, for any $S_{-i}\in \P$, the frequency with which player $i$'s total bid $b_{i}$ exceeds the value of the prize is equal to $\min\{ b_{i1},\ubar{S_{-i}}\}$. In particular, under the F treatment, this equals $\min\{b_{i1},b_{-i1}\}$.
\end{hypothesis}

Focusing on the F treatment, $\min\{b_{i1}, b_{-i1}\}$ represents sunk costs after stage 1 and shouldn't influence second stage bids. Consequently, total bid should exceed the value of the prize with a probability equal to $\min\{b_{i1},b_{-i1}\}$. If the observed behavior reveals a stronger effect of sunk costs, it can be interpreted as evidence in favor of the sunk cost fallacy, as bidders continue to invest more heavily simply because they have already committed significant resources. On the other hand, if the observed behavior reveals a weaker effect, it may reflect a reluctance to bid beyond the prize value, even though doing so would be consistent with equilibrium play.\\

Our last two hypotheses are about the discouragement effect. The first one is about the probability of dropping out, while the second one is about the mean second stage bid.

\begin{hypothesis}[Discouragement effect - 1]
\label{hypo:drop_out}
Under Assumption \ref{ass:CSS_play}, for any $S_{-i}\in \P$, the frequency with which player $i$'s second-stage bid $b_{i2}$ is zero is equal to $|b_{i1} - \ubar{S_{-i}}|$. In particular, for the F treatment, this equals $|b_{i1} - b_{-i1}|$.
\end{hypothesis}

\begin{hypothesis}[Discouragement effect - 2]
\label{hypo:mean}
Under Assumption \ref{ass:CSS_play}, for any $S_{-i}\in \P$, the average second-stage bid $b_{i2}$ of player $i$ is
\begin{align*}
\E[b_{i2}]= \begin{cases} \frac{1}{2}\left[1-(b_{i1}-\ubar{S_{-i}})\right]^2 & \text{if } b_{i1} \geq  \ubar{S_{-i}} \\ \frac{1}{2}\left[1-(b_{i1}-\ubar{S_{-i}})^2\right] & \text{if } b_{i1} \leq  \ubar{S_{-i}} .\end{cases}
\end{align*}
\end{hypothesis}

These hypotheses capture how an increase in the perceived head start (or heterogeneity) is detrimental for bids in the second stage. Specifically, they are more likely to be $0$, and also their mean is lower. Hypotheses \ref{hypo:same_profit} through \ref{hypo:exceed_prize} were pre-registered. Hypotheses \ref{hypo:drop_out} and \ref{hypo:mean} were added after the pre-registration, but are derived from the same model and equilibrium.


\section{Experiment Results}

In this section, we present our findings from the experiment. We normalize the data by dividing all bids and cutoffs by 20 to maintain consistency with the theoretical model.

\subsection{Observations}
\begin{figure}[htbp]
    \centering
    \caption{Bid Distributions}
    \includegraphics[width=\textwidth]{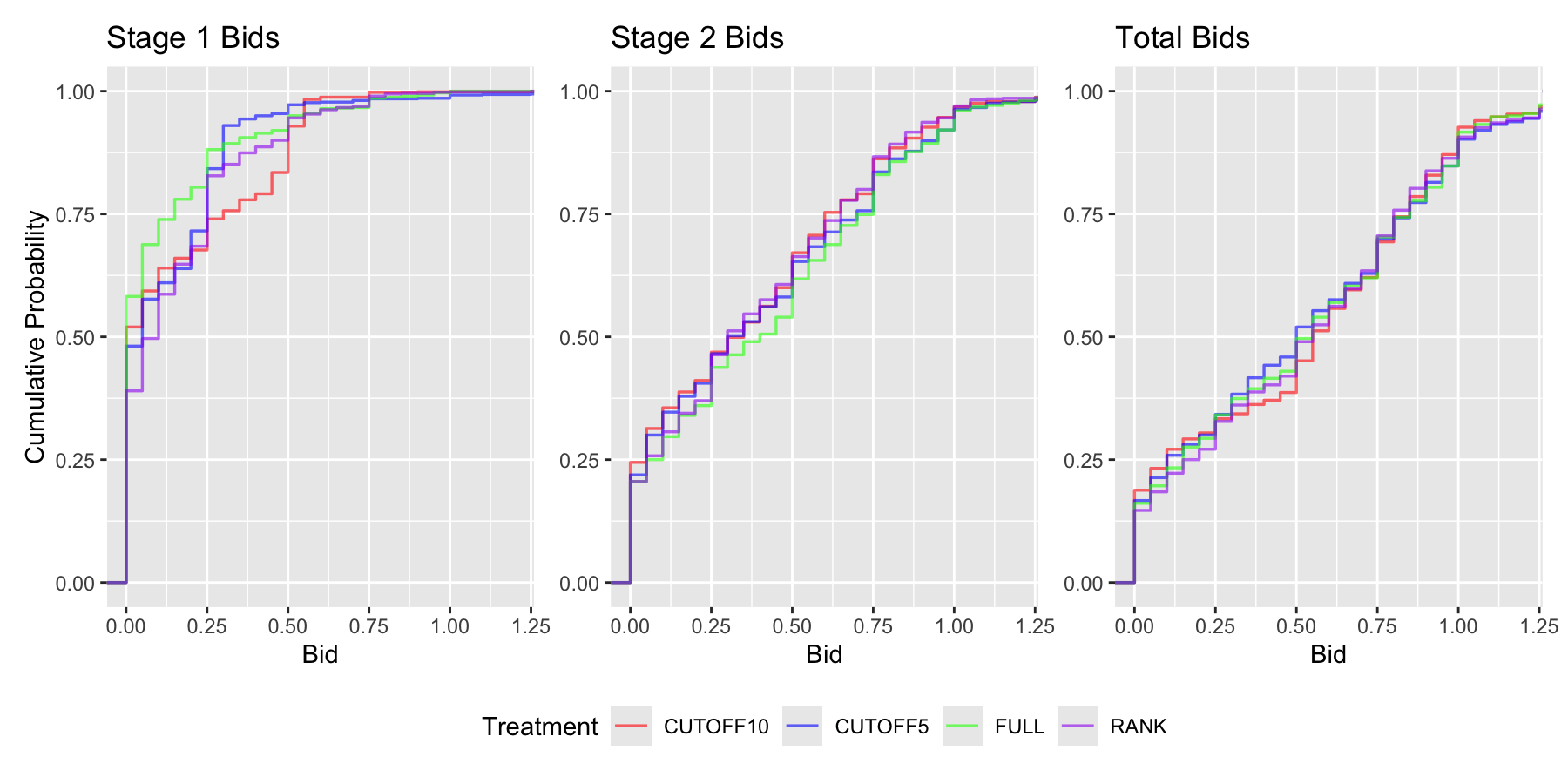}
    \label{fig:bids_cdf}
\end{figure}

We begin with some observations from the empirical CDF's of stage 1 bids, stage 2 bids, and total bids, shown in Figure~\ref{fig:bids_cdf}. \\

The distributions of total bids are broadly similar across treatments, implying a similarity of means, consistent with the irrelevance result. However, there is also evidence of behavior that deviates from equilibrium predictions. The total bid exceeds the prize value roughly 10\% of the time, which should never occur in equilibrium. That said, such bids may result from coordination failures in stage 1, consistent with the sunk-cost effect. In the cutoff treatments, nearly 40\% of stage 1 bids deviate from cheapest signal bidding (i.e., they are not $0$ or $\frac{5}{20}=0.25$ in C5, and not $0$ or $\frac{10}{20}=0.5$ in C10). Moreover, the frequency with which total bid equals the cutoff should be the same as the frequency with which it equals zero, yet total bids of zero are substantially more common than those at the cutoff in either treatment, which indicates a pronounced discouragement effect.  

\subsection{Irrelevance result}

We now discuss Hypotheses \ref{hypo:same_profit} and \ref{hypo:zero_profit}, derived from the irrelevance result in Theorem \ref{thm:irrelevance} and Proposition \ref{prop:rank}.\\

Hypothesis \ref{hypo:same_profit} states that auctioneer profits are equal across treatments. We test this using two-sample t-tests. Table~\ref{tab:hyp1} reports the $t$-statistics, with $p$-values in parentheses, for all six pairwise treatment comparisons. The hypothesis of equal auctioneer profit is not rejected for any of these comparisons.\\

\begin{table}[!h]
\centering
\caption{Test for Hypothesis \ref{hypo:same_profit} (Same profit)}
\centering
\begin{tabular}[t]{lrrr}
\toprule
 & R & C5 & C10 \\
\midrule
F & -0.526 & 0.065 & -0.158 \\
 & (0.60) & (0.95) & (0.87) \\
R & . & 0.560 & 0.369 \\
 & & (0.58) & (0.71) \\
C5 & . & . & -0.213 \\
 & & & (0.83) \\
\bottomrule
\end{tabular}
\label{tab:hyp1}
\end{table}

Hypothesis~\ref{hypo:zero_profit} states that auctioneer profits are zero across treatments. We test this using one-sample $t$-tests with a null hypothesis of zero mean profit. The results are presented in Table~\ref{tab:hyp2}. The $p$-values for all treatments are essentially zero, allowing us to reject the hypothesis of zero auctioneer profits. 

\begin{table}[!h]
\centering
\caption{Test for Hypothesis \ref{hypo:zero_profit} (Zero profit)}
\centering
\begin{tabular}[t]{lrrrrr}
\toprule
Treatment & Mean & 95\% CI Lower & 95\% CI Upper & t-statistic & p-value\\
\midrule
F & 0.072 & 0.034 & 0.110 & 3.74 & 1.96e-4 \\
R & 0.087 & 0.046 & 0.127 & 4.23 & 2.54e-5 \\
C5 & 0.070 & 0.027 & 0.112 & 3.24 & 1.23e-3 \\
C10 & 0.076 & 0.038 & 0.115 & 3.89 & 1.08e-4 \\
\bottomrule
\end{tabular}
\label{tab:hyp2}
\end{table}

In summary, while the auctioneer earns a positive profit, the magnitude of this profit remains similar across treatments, consistent with the irrelevance hypothesis.

\subsection{Second stage bidding}
In this subsection, we discuss Hypotheses \ref{hypo:exceed_prize} through \ref{hypo:mean}, which concern the sunk cost effect and discouragement effect in second stage bidding behavior identified in Proposition \ref{prop:CSS_second_stage}.\\

For these hypotheses, we estimate the coefficients from a series of linear regressions to assess whether these coefficients are consistent with their predicted values. Each regression has a dependent variable $y_i$, which is a function of $b_{i2}$, and takes the following form:

$$
y_{i} = \beta_{0} 
+ \beta_{1} \underbrace{ \min \{ b_{i1}, \underline{S_{-i}} \}}_{\text{Sunk Cost}} 
+ \beta_{2} \underbrace{ |b_{i1}-\underline{S_{-i}}|}_{\text{Head Start}} 
+ \beta_{3} \underbrace{ |b_{i1}-\underline{S_{-i}}|^{2}}_{\text{Head Start$^2$}} 
+ \epsilon_{i}.
$$

For Hypothesis \ref{hypo:exceed_prize}, we use $y_i = 1\{b_i > 1\}$, an indicator for whether the total bid exceeds the prize value. 

For Hypothesis \ref{hypo:drop_out}, we use $y_{i} = 1 \lbrace b_{i2} = 0 \rbrace$, an indicator for whether the stage 2 bid is zero. 

For Hypothesis \ref{hypo:mean}, we use $y_{i} = b_{i2}$.

This specification is chosen because it is the minimal model that nests the equilibrium predictions for each of these dependent variables. Moreover, it captures the two key determinants of stage 2 bidding that we believe are strategically and behaviorally interesting, sunk costs and head starts.\\

We estimate each regression on the F treatment alone, since in this treatment any stage 1 bid satisfies Assumption \ref{ass:CSS_play}. We also estimate using pooled data with the cutoff treatments, while restricting the sample to stage 1 bids that satisfy Assumption \ref{ass:CSS_play}. We also estimate separate regressions for  leaders and laggards after stage 1 bidding. \\

Hypothesis \ref{hypo:exceed_prize} states that if a player enters stage 2 having bid $b_{i1}$ in stage 1 (which is the cheapest bid for some signal), and observes a signal $S_{-i}\in \P$, then its stage 2 bid must be such that the total bid exceeds the value of the prize with probability $\min\{b_{i1},\underline{S_{-i}}\}$, which is its sunk cost. This yields the following hypothesized linear probability model:

$$
1\{b_{i}>1\} =  \text{Sunk Cost}_i + \epsilon_{i}. 
$$

Thus, the probability that the total bid exceeds the prize value is increasing in sunk costs, and independent of the head start. Behaviorally however, players who have already invested more in stage 1 may bid more aggressively in stage 2, in line with the sunk cost fallacy. Alternatively, players may be reluctant to submit total bids above the prize value, since doing so guarantees a negative final payoff. 

\begin{table}[htbp] \centering 
  \caption{Test for Hypothesis \ref{hypo:exceed_prize} (Exceed prize value)} 
  \label{tab:over_reg} 
\begin{tabular}{@{\extracolsep{5pt}}lcccccc} 
\\[-1.8ex]\hline 
\hline \\[-1.8ex] 
& \multicolumn{3}{c}{\textbf{Treatment: F}} & \multicolumn{3}{c}{\textbf{Treatments: F, C5, C10}} \\
\cmidrule(lr){2-4} \cmidrule(lr){5-7}
 & Pooled & Leader & Laggard & Pooled & Leader & Laggard \\ 
\\[-1.8ex] & (1) & (2) & (3) & (4) & (5) & (6)\\ 
\hline \\[-1.8ex] 
 Sunk Cost & 0.87 & 1.06 & 0.77 & 0.50 & 0.46 & 0.33 \\ 
  & (0.11) & (0.14) & (0.11) & (0.06) & (0.07) & (0.06) \\ 
  & & & & & & \\ 
 Head Start & 0.16 & 0.28 & $-$0.02 & 0.10 & 0.16 & 0.06 \\ 
  & (0.11) & (0.15) & (0.12) & (0.07) & (0.09) & (0.07) \\ 
  & & & & & & \\ 
 Head Start$^2$ & 0.14 & 0.20 & 0.13 & 0.20 & 0.32 & 0.03 \\ 
  & (0.15) & (0.21) & (0.16) & (0.10) & (0.14) & (0.12) \\ 
  & & & & & & \\ 
 Constant & 0.03 & 0.03 & 0.03 & 0.02 & 0.02 & 0.02 \\ 
  & (0.01) & (0.01) & (0.01) & (0.01) & (0.01) & (0.01) \\ 
  & & & & & & \\ 
\hline \\[-1.8ex] 
Observations & 900 & 600 & 600 & 2,000 & 1,507 & 1,556 \\ 
R$^{2}$ & 0.11 & 0.16 & 0.08 & 0.06 & 0.09 & 0.02 \\
\hline 
\hline \\[-1.8ex] 
\textit{Note:} & \multicolumn{6}{r}{Standard errors in parentheses.} \\ 
\end{tabular} 
\end{table}

The linear probability regression estimates for Hypothesis \ref{hypo:exceed_prize} are reported in Table \ref{tab:over_reg}. In the F treatment, the coefficient on sunk cost is not statistically different from $1$, though it declines significantly when data from the C5 and C10 treatments are included. The head start coefficients are not statistically different from zero, and while the constant term is positive and significant, its magnitude is small. We interpret these findings as broadly consistent with the equilibrium predictions.\\

Hypothesis \ref{hypo:drop_out} states that if a player enters stage 2 having bid $b_{i1}$ in stage 1 (which is the cheapest bid for some signal), and observe a signal $S_{-i}\in \P$, then its stage 2 bid must be zero with probability $ |b_{i1} - \ubar{S_{-i}}|$, which is the head start. This yields the following hypothesized linear probability model:

$$
1\{b_{i2} = 0\} =  \text{Head Start}_i + \epsilon_{i} 
$$

Thus, the probability that a player bids zero in stage 2 is increasing in head start, and independent of the sunk costs. This reflects the discouragement effect widely studied in the contest literature: as contest becomes unfair, incentives to exert effort are weakened. \\

\begin{table}[!htbp] \centering 
  \caption{Test for Hypothesis \ref{hypo:drop_out} (Zero stage 2 bid)} 
  \label{tab:zero_reg} 
\begin{tabular}{@{\extracolsep{5pt}}lcccccc} 
\\[-1.8ex]\hline 
\hline \\[-1.8ex] 
& \multicolumn{3}{c}{\textbf{Treatment: F}} & \multicolumn{3}{c}{\textbf{Treatments: F, C5, C10}} \\
\cmidrule(lr){2-4} \cmidrule(lr){5-7}
 & Pooled & Leader & Laggard & Pooled & Leader & Laggard \\ 
\\[-1.8ex] & (1) & (2) & (3) & (4) & (5) & (6)\\ 
\hline \\[-1.8ex] 
 Sunk Cost & $-$0.51 & $-$0.41 & $-$0.66 & $-$0.57 & $-$0.48 & $-$0.61 \\ 
  & (0.16) & (0.19) & (0.22) & (0.12) & (0.13) & (0.14) \\ 
  & & & & & & \\ 
 Head Start & 0.35 & $-$0.30 & 0.83 & 0.26 & $-$0.60 & 0.89 \\ 
  & (0.17) & (0.20) & (0.23) & (0.13) & (0.16) & (0.17) \\ 
  & & & & & & \\ 
 Head Start$^2$ & $-$0.05 & 0.50 & $-$0.45 & $-$0.04 & 0.83 & $-$0.62 \\ 
  & (0.23) & (0.28) & (0.32) & (0.20) & (0.26) & (0.28) \\ 
  & & & & & & \\ 
 Constant & 0.16 & 0.17 & 0.18 & 0.25 & 0.25 & 0.26 \\ 
  & (0.02) & (0.02) & (0.02) & (0.01) & (0.01) & (0.01) \\ 
  & & & & & & \\ 
\hline \\[-1.8ex] 
Observations & 900 & 600 & 600 & 2,000 & 1,507 & 1,556 \\ 
R$^{2}$ & 0.04 & 0.01 & 0.07 & 0.02 & 0.02 & 0.06 \\ 
\hline 
\hline \\[-1.8ex] 
\textit{Note:} & \multicolumn{6}{r}{Standard errors in parentheses.} \\ 
\end{tabular} 
\end{table} 

The linear probability regression estimates for Hypothesis \ref{hypo:drop_out} are reported in Table \ref{tab:zero_reg}. Sunk costs have a significant negative effect on the likelihood of bidding $0$ in the second stage, even though it should have no influence in equilibrium. Moreover, the coefficient on head start is significantly below one, while the constant term is positive and large. These findings indicate a departure from equilibrium behavior and may be interpreted as evidence consistent with the sunk-cost fallacy.\\

Finally, Hypothesis \ref{hypo:mean} states that if a player enters stage 2 having bid $b_{i1}$ in stage 1 (which is the cheapest bid for some signal), and observe a signal $S_{-i}\in \P$, then its mean stage 2 bid must take the form in Proposition \ref{prop:CSS_second_stage}, which yields the following hypothesized linear model for the discouragement effect:

\begin{align*}
b_{i2} &= \begin{cases} \frac{1}{2}-\text{Head Start}_i + \frac{1}{2} \text{Head Start}_i^2 +\epsilon_i & \text{if } b_{i1} \geq  \ubar{S_{-i}} \\ \frac{1}{2}-\frac{1}{2}\text{Head Start}_i^2 +\epsilon_i & \text{if } b_{i1} \leq  \ubar{S_{-i}} .\end{cases} 
\end{align*}

Thus, the mean stage 2 bid decreases with the head start (for both leaders and laggards), another manifestation of the discouragement effect, and is independent of sunk costs.\\

\begin{table}[!htbp] \centering 
  \caption{Test for Hypothesis \ref{hypo:mean} (Mean stage 2 bid)} 
  \label{tab:bid2_reg} 
\begin{tabular}{@{\extracolsep{5pt}}lcccc} 
\\[-1.8ex]\hline 
\hline \\[-1.8ex] 
 & \multicolumn{2}{c}{\textbf{F}} & \multicolumn{2}{c}{\textbf{F, C5, C10}} \\
\cmidrule(lr){2-3} \cmidrule(lr){4-5}
 & Leader & Laggard & Leader & Laggard \\ 
\\[-1.8ex] & (1) & (2) & (3) & (4)\\ 
\hline \\[-1.8ex] 
 Sunk Cost & 0.06 & 0.52 & 0.15 & 0.31 \\ 
  & (0.19) & (0.21) & (0.12) & (0.13) \\ 
  & & & & \\ 
 Head Start & 0.16 & 0.22 & 0.10 & 0.19 \\ 
  & (0.20) & (0.22) & (0.14) & (0.15) \\ 
  & & & & \\ 
 Head Start$^2$ & $-$0.47 & $-$0.56 & $-$0.37 & $-$0.44 \\ 
  & (0.28) & (0.30) & (0.23) & (0.24) \\ 
  & & & & \\ 
 Constant & 0.44 & 0.42 & 0.38 & 0.37 \\ 
  & (0.02) & (0.02) & (0.01) & (0.01) \\ 
  & & & & \\ 
\hline \\[-1.8ex] 
Observations & 600 & 600 & 1,507 & 1,556 \\ 
R$^{2}$ & 0.01 & 0.02 & 0.01 & 0.01 \\ 
\hline 
\hline \\[-1.8ex] 
\end{tabular} 
\end{table} 

The linear regression estimates for Hypothesis \ref{hypo:mean} are reported in Table \ref{tab:bid2_reg}. Sunk costs have a positive effect on stage 2 bids, though the coefficients are insignificant for leaders. The coefficients on head start and its square imply that the marginal effect of increasing head start on the stage 2 bid is positive for small values but becomes negative thereafter. The constant term is positive but significantly below $\frac{1}{2}$. Overall, these results provide mixed evidence: they reveal a discouragement effect of increasing head start, while also suggesting the presence of a sunk-cost fallacy that is more pronounced among laggards than leaders.\\
  
Taken together, the regression results suggest that the bidding behavior deviate from equilibrium (or CSS), and exhibits the following behavioral patterns. Higher sunk costs are linked to a lower probability of dropping out and to higher stage 2 bids, consistent with the sunk-cost fallacy. Nonetheless, this effect is somewhat moderated by the prize value, as total bids do not exceed the prize more frequently than predicted. Higher head starts do seem to be associated with lower stage 2 bids, though this discouragement effect appears to kick in only when the head start is not too small. Moreover, this effect is not as pronounced as the equilibrium predicts. 
\section{Conclusion}

In this paper, we study how feedback policies affect investment behavior in a two-stage, two-player all-pay auction. We introduce the notions of Cheapest Signal Beliefs and Cheapest Signal Strategies and show that they characterize the set of (stage 1 pure) sequential equilibrium outcomes. In any equilibrium, stage 1 bids are such that one player bids zero while the other chooses the cheapest bid consistent with some signal, and stage 2 bidding mimics the unique (mixed) Nash equilibrium of an all-pay auction with exogenous head starts. This characterization yields an irrelevance result: equilibrium payoffs for both players and the auctioneer’s profits are zero, regardless of the feedback policy. Further, since stage 1 bids induce sunk costs and head starts, we examine how these features influence stage 2 bidding, focusing on the sunk-cost effect (prior investments should not affect subsequent bids), and the discouragement effect (greater disparities in early bids reduce later investment).\\

We conducted a laboratory experiment to test these predictions. We used four feedback policy treatments, a full feedback treatment, a rank treatment, and two cutoff treatments. While the auctioneer obtains positive profits across treatments, the difference in magnitudes is statistically insignificant and we fail to reject the hypothesis of no treatment effect on total bids. For stage 2 bidding, we find that the bids are increasing in sunk costs, in line with the sunk cost fallacy, though the effect is somewhat guarded by the prize value. The discouragement effect of higher head starts kicks in when it is not too small, and is not as large as predicted.\\

Our analysis suggests several promising avenues for future research. To begin, it would be interesting to study whether the effect of sunk costs and head start are robust to being exogenously fixed, or does the fact that the players actually chose the stage 1 bids matter for how they bid in the stage 2. Additionally, since bids in the two stages are perfect substitutes, there isn't really a benefit to bidding in stage 1, except perhaps to deter the other player. It would be interesting to study variations of the model with convex costs, or private types, where stage 1 bids would potentially have greater significance.

\newpage
\bibliographystyle{ecta}

\bibliography{refs}

\end{document}